\documentclass[aps,twocolumn,superscriptaddress,prl]{revtex4-2} \usepackage{amssymb,amsmath,amsthm,amsfonts,amsbsy,mathrsfs}
\usepackage{xcolor,hyperref}
\hypersetup{
   colorlinks,
   linkcolor={blue!50!black},
   citecolor={blue!50!black},
   urlcolor={blue!80!black}
}
\usepackage{graphicx}
\usepackage{color}
\usepackage{bm}
\usepackage[english]{babel}

\newcommand{\xe}{\xi_{\rm e}}

\newcommand{\obp}{\omega_{\rm bp}}
\newcommand{\Dql}{D_{\rm loc}}

\usepackage{physics}
\begin{document}

\title{
Unifying description of the vibrational anomalies of amorphous materials}

\author{Shivam Mahajan}
\affiliation{Division of Physics and Applied Physics, School of Physical and
Mathematical Sciences, Nanyang Technological University, Singapore}

\author{Massimo Pica Ciamarra}
\email{massimo@ntu.edu.sg}
\affiliation{Division of Physics and Applied Physics, School of Physical and
Mathematical Sciences, Nanyang Technological University, Singapore}

\date{\today}

\begin{abstract}
The vibrational density of states $D(\omega)$ of solids controls their thermal and transport properties. 
In crystals, the low-frequency modes are extended phonons distributed in frequency according to Debye's law, $D(\omega) \propto \omega^2$.
In amorphous solids, phonons are damped,
and at low frequency $D(\omega)$ comprises extended modes in excess over Debye's prediction, leading to the so-called boson peak in $D(\omega)/\omega^2$ at $\obp$, and quasi-localized (QLMs) ones.
Here we show that boson peak and phonon attenuation in the Rayleigh scattering regime are related, as suggested by correlated fluctuating elasticity theory (corr-FET), and that amorphous materials can be described as homogeneous isotropic elastic media punctuated by QLMs acting as elastic heterogeneities. 
Our numerical results resolve the conflict between theoretical approaches attributing amorphous solids' vibrational anomalies to elastic disorder and localized defects. 
\end{abstract}
\maketitle

The distribution in the frequency, $\omega$, of the vibrational modes of solids, or density of states (DOS), is a fundamental material property controlling, e.g., their specific heat and thermal conductivity~\cite{kittel1996, ZellerAmorphous}.
At small frequencies, crystals' DOS is populated by phonons (plane waves) and follows Debye's law, $D(\omega) = A_D \omega^2$.
The vibrational properties of amorphous materials deviate from that of crystals in several aspects.
First, the reduced density of states $D(\omega)/\omega^2$ exhibits a peak at the Boson peak frequency, $\obp$, in the terahertz regime for molecular solids.
The boson peak reveals an excess of modes over Debye's prediction. Competing theories have attributed this anomaly to elastic disorder 
~\cite{SchirmacherPRL, SchirmacherEPL2006, Marruzzo2013a,Elliott1992,Graebner1986, Kawahara2011}, localized harmonic~\cite{Phillips1981, Maurer2004} or anharmonic vibrations~\cite{Ruffle2008, Buchenau1992}, anharmonic effects~\cite{Baggioli2019,yang2021giant}, broadening of the lowest van-Hove singularity of the transverse phonon branch~\cite{Taraskin2001, Chumakov2011}.
Second, the low-frequency DOS of amorphous solids is the superposition~\cite{Mizuno2014, kapteijns2018universal, richard2020universality, rainone2020pinching} of extended modes complying to Debye's prediction, $D(\omega) = A_D\omega^2$, and of quasi-localized modes (QLMs) distributed in frequency as  $\Dql(\omega) = A_4 \omega^4$.
Finally, in amorphous solids, the extended low-frequency modes are not phonons.
Rather, phonons attenuate while propagating with a frequency-dependent rate, $\Gamma(\omega)$.
In the absence of temperature induced anharmonic effects~\cite{Mizuno2020}, phonons' attenuation rate $\Gamma$ crossovers from a Rayleigh scattering regime~\cite{Strutt1903}, $\Gamma \propto \omega^{4}$, to a disordered-broadening regime, $\Gamma \propto \omega^{2}$~\cite{Masciovecchio2006,Baldi2010,monaco2009breakdown,Ruta2012,Baldi2013}, as the phonon frequency increases, as observed in recent studies~\cite{Moriel2019, wang2019sound, kapteijns2021elastic}.

Since the vibrational anomalies of amorphous solids occur in different frequency regimes, it is not clear that there should be a relation between them~\cite{Wang2019,wang2019sound}. 
However, there are indications suggesting such a relation.
For instance, numerical results indicate a correlation between $A_4$ and $\obp$~\cite{Mizuno2017, Wang2019}.
Furthermore, fluctuating elasticity theory~\cite{SchirmacherPRL, Schirmacher1998, SchirmacherEPL2006, Marruzzo2013a} (FET), in its extended version incorporating an elastic disorder correlation length $\xe$~\cite{Schirmacher2010, Schirmacher2011} (corr-FET) suggests a correlation between $\obp$ and the attenuation rate of sound waves in the Rayleigh scattering regime,  $\Gamma/ \obp \propto \gamma (\omega/\obp)^4$. 
Here, $\gamma$ is a disorder parameter controlling the scaling of the fluctuations $\sigma^2_\mu(N)$ of the shear modulus on the coarse-grained size~\cite{SchirmacherPRL, Schirmacher1998, SchirmacherEPL2006, Marruzzo2013a}, $\sigma^2(N)/\mu^2 = \gamma/N$, $\mu$ being the average modulus, and $\obp = c_s/\xe$ with $c_s$ the sound velocity of transverse waves.
While sound attenuation appears to correlate with the fluctuations of the elastic moduli~\cite{Mizuno2014,kapteijns2021elastic}, the validity of corr-FET is debated. 
It has been suggested, for instance, that corr-FET is only qualitatively accurate~\cite{wang2019sound,Caroli2019} or that the frequency controlling sound attenuation is not $\obp$ but
$\omega_0 =c_s/a_0$, with $a_0 = \rho^{-1/d}$ and $\rho$ the number density~\cite{kapteijns2021elastic}. 
Henceforth, it is still unclear if boson peak, quasi-localized modes and sound attenuation are related.

Here, we introduce and verify via extensive numerical simulations a simple picture relating amorphous solids' vibrational anomalies.
First, we validate corr-FET and its proposed connection between boson peak, elastic heterogeneities and sound attenuation.
Then, we show that low-frequency corr-FET' predictions emerge from the mechanical model introduced by Rayleigh in his seminal work~\cite{Strutt1903},
a homogeneous elastic continuum of shear modulus $\mu_0$ punctuated by defects with shear modulus $\mu_0+\delta\mu_d$, provided that the defects have linear size $\xi_d \propto \xe$, constant number density $n$, and that $\delta\mu_d \propto \mu_0$.
Finally, we demonstrate that QLMs satisfy these constraints. 
Our results clarify that the low-frequency vibrational properties of amorphous solids are those of an elastic continuum punctuated by quasi-localized vibrational modes. 
Hence, our work establishes a relation between the different vibrational anomalies of amorphous solids and resolves the contrast between theoretical models attributing the boson peak anomaly to elastic disorder and localized defects. 

We resort to numerical simulations to investigate vibrational properties and attenuation rate of model amorphous materials, focusing on systems of particles interacting via an LJ-like potential $V(r,x_c)$. Here, $x_c$ is a parameter $x_c$ setting the extension of the attractive well~\cite{dauchotPotential}, which vanishes at $x_c \sigma$, controlling the relaxation dynamics~\cite{chattoraj2020role} and the mechanical response~\cite{dauchotPotential, gonzalez2020mechanical, gonzalez2020mechanical2, Zheng2021}. 
We follow the model of Ref.~\cite{chattoraj2020role}.
We simulate systems with a varying number of particles $N$, up to $N = 8192000$, in a cubic box with periodic boundary conditions, at fixed number density $\rho = 1.07$.
We generate amorphous solid configurations by minimizing, via conjugate gradient, the energy of systems in thermal equilibrium in the NVT ensemble at $T=4.0\epsilon$, above the glass transition temperature for the considered $x_c$ values~\cite{chattoraj2020role}.

\begin{figure}[t!]
 \centering
 \includegraphics[angle=0,width=0.48\textwidth]{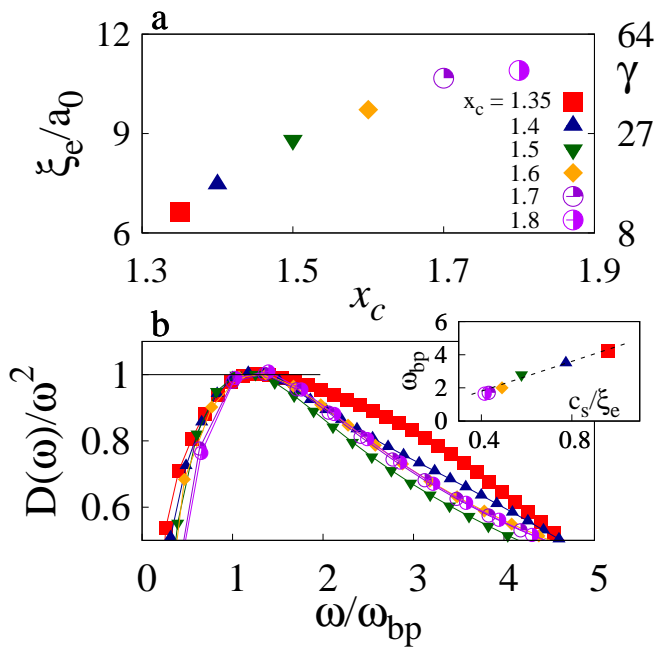}
 \caption{
(a) Dependence of elastic disorder correlation length $\xe \propto \gamma^{1/3}$ and disorder parameter $\gamma$ on $x_c$, a parameter controlling the extension of the attractive well. 
Error bars are smaller than the symbol size.
(b) Reduced $D(\omega)$, normalized by its maximum value, as a function of $\omega/\obp$.
We found $\obp \simeq 4.5 c_s/\xe$ (inset).
Data are Bezier-smoothed for clarity.
\label{fig:ne}
}
\end{figure}

Corr-FET predicts~\cite{SchirmacherPRL, Schirmacher1998, SchirmacherEPL2006, Marruzzo2013a} that the boson peak frequency, $\obp\simeq c_s/\xe$, and sound attenuation in the Rayleigh scattering regime, $\frac{\Gamma(\omega)}{\obp} \propto \gamma \left(\frac{\omega}{\obp}\right)^4$, so that
\begin{equation}
\Gamma \frac{\omega_0^3}{\omega^4} \propto \gamma^a
\propto \left(\frac{\omega_0}{\obp}\right)^{3a}
{\rm~with~} a = 2. 
\label{eq:Gcf}
\end{equation}
Conversely, according to early FET theory~\cite{SchirmacherPRL, SchirmacherEPL2006, Marruzzo2013a, richard2020universality}, $a = 1$. 
To validate this theory, we determine $\xe$ and $\gamma$ investigating the system and coarse-grained size dependence of the elastic properties, finding consistent results.
We follow standard approaches~\cite{Tsamados2009,mahajan2021emergence,gonzalez2020mechanical} and provide details in the Supplementary Material~\cite{SM}.
We find that a single length scale characterizes the dependence the shear modulus fluctuations on the system size, so that $\gamma$ is a non-dimensional measure of the correlation volume, $\gamma \propto \xe^3$.
In Fig.~\ref{fig:ne}(a), we observe the elastic length scale $\xe$, or equivalently $\gamma$, to decrease with the attraction range $x_c$.

The estimation of $\xe$, $\gamma$ and $c_s = \sqrt{\mu/(m\rho)}$ allows us to validate if, as predicted by corr-FET, the boson peak frequency scales as $\obp \propto c_s/\xe$.
To check this prediction, we evaluate $D(\omega)$ via the Fourier transform of the velocity autocorrelation function of N = 256000 particle systems, and $\obp$ via the scaling collapse of Fig.~\ref{fig:ne}(b). 
In the inset, we show that the boson peak frequency is proportional to $c_s/\xe$. Hence, corr-FET correctly predicts the relation between boson peak and elastic heterogeneities.

We now consider if corr-FET correctly predicts sound attenuation rate in the Rayleigh scattering regime.
To evaluate the attenuation rate, we excite~\cite{Gelin2016, Mizuno2018} a transverse acoustic wave with wave vector $\boldsymbol{\kappa}$ in which two among $\kappa_x$, $\kappa_y$ and $\kappa_z$ are zero. 
We then follow the system's time evolution in the linear response regime to evaluate the velocity auto-correlation function, which we average over 30 phonons from independent samples for $N \leq 512000$, and over 15 phonons for $N \geq 512000$.
A subsequent fit of this averaged velocity autocorrelation function to $\cos(\omega t)e^{-\Gamma t/2}$ allows extracting attenuation rate $\Gamma$ and frequency $\omega$ as a function of $\kappa$.

\begin{figure}[t!]
\centering
\includegraphics[angle=0,width=0.48\textwidth]{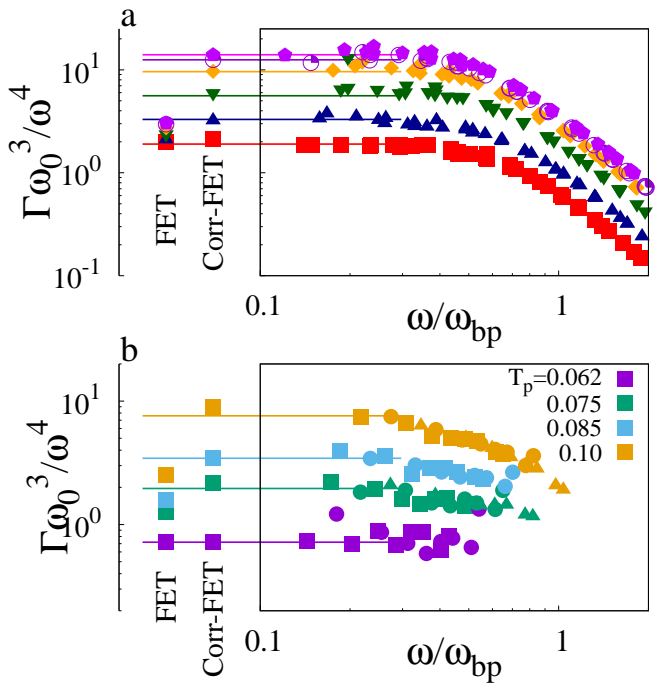}
\caption{
(a) The frequency dependence of the scaled attenuation rate is consistent with corr-FET as concern the limiting low-frequency value, $\propto \gamma^2$. We combine data for N = 32k, 64k, 256k, 512k, 2048k and 8192k.
Symbols are as in Fig.~\ref{fig:ne}. 
(b) Analogous results are obtained investigating the scaled sound attenuation rate of amorphous solid configurations prepared minimizing the energy of ultrastable liquids in equilibrium at temperature $T_p$, below the mode coupling one.
Data are form Ref.~\cite{wang2019sound,Shakerpoor2020}, to which we refer for further details. 
Symbols identify the system size: 192k (squares), 96k (circles), 48k (triangles).
\label{fig:attPar}
}
\end{figure}

The normalized attenuation rate $\Gamma \omega_0^3/\omega^4$ attains a constant value at low frequency, demonstrating the existence of a well-defined Rayleigh scattering regime, as illustrated in Fig.~\ref{fig:attPar}(a). This finding~\cite{Moriel2019, wang2019sound, kapteijns2021elastic} demonstrates that, in this regime, anisotropic long-range spatial correlations in the elastic moduli~\cite{Lemaitre2015,Shivam2021} do not influence sound damping~\cite{Gelin2016,Cui2020}.
We test FET and corr-FET predictions, Eq.~\ref{eq:Gcf}, by plotting $\alpha\gamma$ and $\beta\gamma^2$, with $\alpha$ and $\beta$ constants and $\gamma$ as in Fig.~\ref{fig:ne}(a).
Corr-FET correctly predicts the relation between sound attenuation and boson peak frequency.
Furthermore, Fig.~\ref{fig:attPar}(a) indicates that Rayleigh's scattering regime sets in at a frequency smaller but close to $\obp$, confirming another corr-FET prediction.

Previous works did not support corr-FET.
Refs.~\cite{Shakerpoor2020,wang2019sound} tested it by measuring the fluctuations of coarse-grained elastic constants defined via the so-called ``fully local'' approach~\cite{Mizuno2013}. 
We speculate this approach leads to unreliable results as it fails to recover self-averaging~\cite{Tsamados2009}.
Ref.~\cite{richard2020universality} supported the validity of FET, rather than of corr-FET, studying sound attenuation and elastic properties as a function of a parameter artificially affecting the pre-stress contribution to the dynamical matrix of a given system.
We suspect this approach breaks the relation $\gamma \propto \xe^3$, leading to changes in $\gamma$ at constant $\xe$, but the matter deserves further investigation.

To further support our findings, we consider that Eq.~\ref{eq:Gcf} can tested without the direct measurement of the disorder parameter, but rather inferring it from measurements of the boson peak frequency.
We exploit this result to validate corr-FET against numerical data for the boson peak frequency~\cite{wang2019sound} and sound attenuation~\cite{Shakerpoor2020} of ultrastable glasses. 
While it would be interesting to investigate larger system sizes to access the Rayleigh scattering regime at all temperatures unambiguously, the result of this investigation further support the validity of corr-FET, as we illustrate in Fig.~\ref{fig:attPar}(b).

Fluctuating elasticity theory does not make any reference to the existence of defects.
On the contrary, it considers that, in an amorphous material, ``it is difficult to distinguish between `host' and `defect' ''~\cite{Maurer2004}.
Yet, the analysis of the vibrational properties of amorphous materials revealed the existence of QLMs, extended soft mechanical regions that act as structural defects controlling the mechanical response under shear~\cite{Manningb} and, possibly, the relaxation dynamics of supercooled liquids~\cite{Widmer-Cooper2008}. 
Hence, there could be a relation between FET and QLMs. 

We establish this relation within Rayleigh's elastic model~\cite{Strutt1903}, an elastic continuum with shear modulus $\mu_0$ punctuated by $n$ defects per unit volume, each defect being a region of linear size $\xi_d$ with shear modulus $\mu_0 + \delta\mu_d$.
Within this model, FET disorder parameter~\cite{SM} results
\begin{equation}
\gamma \propto (na_0^3) \left( \frac{\xi_d}{a_0}\right)^3\frac{\delta\mu^2_d}{\mu_0^2}
\label{eq:gdefect}
\end{equation}
and the boson peak frequency is $\obp \propto c_s/\xi_d$, so that Rayleigh's seminal result for the attenuation rate~\cite{Strutt1903} of low-frequency phonons, $\Gamma \propto \left(\frac{\delta\mu_d}{\mu_0}\right)^2 \xi_d^6 \omega^4$, can be expressed as 
$\Gamma(\omega) \frac{\xi_d}{c_s} \propto \gamma \left(\frac{\omega \xi_d}{c_s}\right)^4$.

Corr-FET and the defect model are consistent in their predictions for the boson peak frequency if 
\begin{equation}
\xi_d \propto \xe \label{eq:xixe}.
\end{equation}
If this relation holds, the models are consistent in their predictions for the attenuation rate if Eq.~\ref{eq:gdefect} is satisfied, or equivalently, given Eq.~\ref{eq:xixe}, if $n \delta\mu^2/\mu_0^2 = {\rm const}$. This occurs, e.g., if 
\begin{eqnarray}
n &=& {\rm const} \label{eq:n} \\
\delta\mu_d &\propto& \mu_0. \label{eq:sig}
\end{eqnarray}
We now show that QLMs satisfy Eqs.~\ref{eq:xixe},\ref{eq:n},\ref{eq:sig}.

\begin{figure}[t!]
 \centering
 \includegraphics[angle=0,width=0.45\textwidth]{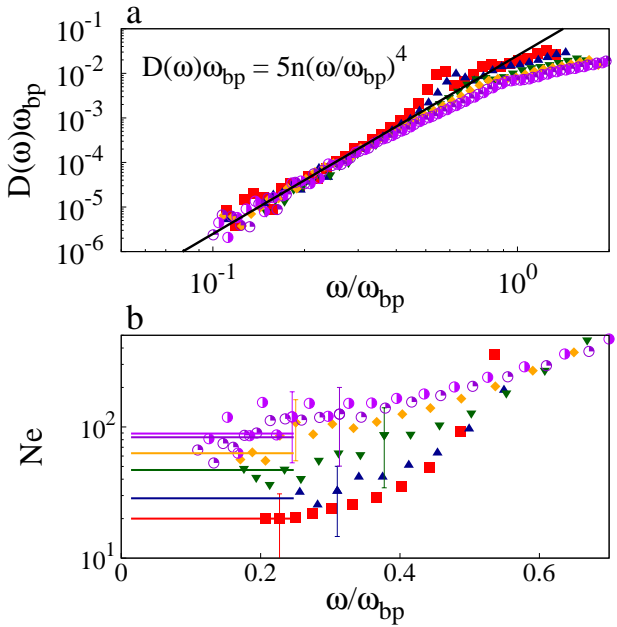}
\caption{
(a) Scaling of the low-frequency density of states.
The full line corresponds to $0.0025 (\omega/\obp)^4$.
(b) QLMs volume estimated by the low-frequency limit of $Ne$ with $e$ the participation ratio and $N$ the system size.
Representative error bars are shown.
The full lines correspond to $\xe^3 \propto \gamma$.
For both panels $N = 4000$ and data are averaged over at least $10^4$ realizations.
Symbols indicate different $x_c$ values as in Fig.~\ref{fig:ne}.
\label{fig:comparexd}
}
\end{figure}

We valide Eq.~\ref{eq:n} investigating QLMs' density of states to estimate their number density, $n$.
We determine the $D_{\rm loc}(\omega)$ via the direct diagonalization of the Hessian of a small systems, $N=4000$, to lift the minimum phonon frequency, $\propto c_s/N^{1/3}$, and make the low-frequency spectrum predominantly populated by localized modes.
We average our results over at least $10^4$ independent realizations.
Fig.~\ref{fig:comparexd}(a) shows that data for different potentials collapse when $D_{\rm loc}$ is non-dimensionalised resorting to the boson peak frequency.
Assuming that $\obp$ is the maximum QLMs' frequency, this result indicates that
\begin{equation}
D_{\rm loc}(\omega) = A_4\omega^4 = \frac{5n}{\obp}\left(\frac{\omega}{\obp}\right)^4,  ~~~\omega <\obp,
\label{eq:dloc}
\end{equation}
where $n = \int_0^{\obp} D(\omega)d\omega \simeq 0.005$ is the {\it constant} density of vibrational modes~\cite{gonzalez2020mechanical}.
Interestingly, Ref.~\cite{Wang2019} reported $A_4^{-1/5}/\obp = (5n)^{-1/5} \simeq 2.1$ (see their Fig. 6), in quantitative agreement with our result, investigating a different system. 
This suggests $n$ might be a universal constant.
We leave to the future the investigation of this intriguing question.

We verify Eq.~\ref{eq:xixe}, which is supported by previous investigations~\cite{rainone2020pinching,gonzalez2020mechanical2}, evaluating QLMs size via the mode participation ratio $e = \frac{1}{N} \left[\sum_{i=1}^N (\vec {u}_i \cdot \vec{u}_i)^2 \right]^{-1}$, where $\vec{u}_i$ is the displacement vector of particle $i$ in the considered mode.
The participation ratio is $\order{1}$ for extended modes, and $\order{1/N}$ for localized ones.
Hence, $Ne$ estimates the number of particles involved in the mode, and we expect $\xi_d^3 \propto \gamma \propto \lim_{\omega \to 0} Ne(\omega)$ if Eq.~\ref{eq:xixe} holds.
While our $Ne$ data are noisy, despite our significant statistics, they are indeed compatible with this expected scenario as we demonstrate in Fig.\ref{fig:comparexd}b, where full lines correspond to $a\gamma(x_c)$, with $a$ constant. 

\begin{figure}[t!]
 \centering
 \includegraphics[angle=0,width=0.45\textwidth]{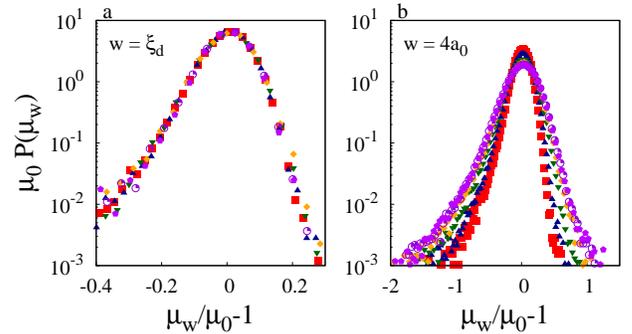}
\caption{
Probability distribution of the local shear modulus $\mu_w/\mu_0-1$ coarse grained over a length scale $w$, $\mu_0$ being the average modulus.
In (a) the coarse graining length scale equals the defect size, $w = \xi_d = \xi_e$, whose $x_c$ dependence is in Fig.~\ref{fig:ne}a. 
In panel (b), $w = 4a_0$.
For each $x_c$, results are obtaining averaging over 50 independent N = 256000 particles configurations.
Symbols are as in Fig.~\ref{fig:ne}.
\label{fig:Gdist}
}
\end{figure}

Since QLMs correspond to soft mechanical regions, we assume that their typical shear modulus is encoded in the left-tail of the distribution of the shear modulus coarse-grained at the QLMs' size $\xi_d$, $P(\mu_{\xi_d})$.
Hence, if Eq.~\ref{eq:sig} holds, the left tails of distributions corresponding to different $x_c$ should collapse, when the distributions are plotted versus $\mu_{\xi_d}/\mu_0-1$, with $\mu_0$ the average shear modulus.
To check this prediction, we associate to each particle a shear modulus, taking into account the non-affine contribution, and coarse-grained it at different length scales~\cite{SM}.
Fig.~\ref{fig:Gdist}a shows that the left tails of the distribution of the shear modulus coarse grained at $w = \xi_d$ collapse - indeed, the whole distribution does it.
We remark that this collapse is not trivial as it occurs at a coarse-graining length scale at which the distributions are far from being Gaussian. 
Indeed, the collapse does not occur at smaller coarse-graining length scale, as we illustrate in Fig.~\ref{fig:Gdist}b where we fix, as an example, $w = 4a_0$.

Overall, these results show that QLMs satisfy Eqs.~\ref{eq:xixe},\ref{eq:n}, \ref{eq:sig} and demonstrate that corr-FET' predictions are recovered within a defect picture, if defects are identified with the QLMs. 
Hence, elastic disorder and defect based interpretations of the anomalous vibrational properties of amorphous materials are intimately related rather than contrasting.
In the defect picture, the density of states of amorphous materials is approximated by
\begin{equation}
D(\omega) = n \frac{\omega^4}{\obp^5} \theta(\obp-\omega) + A_D\frac{\omega^2}{\omega_D^3}\theta(\omega_D-\omega)
\end{equation}
with $\theta(x)$ the Heaviside step function, $n$ weakly system dependent if not constant, and $A_D$ fixed by the normalization constraint.
The characteristic QLMs size determines the boson peak.
Hence, this is not related to the first van Hove singularity of transverse waves, in agreement with a recent finding~\cite{zhang2021disorderinduced}.

QLMs also dominate sound attenuation in the Rayleigh scattering regime, where this is strongly dependent on the defect size, $\Gamma \propto \xi_d^6$, and the deviation of the elastic properties of the defects from the average, $\Gamma \propto \delta \mu_d^2$.
We speculate this occurs because, while in amorphous materials it is undoubtedly difficult to disentangle host and defects, QLMs stands out as the largest and softest elastic heterogeneities from the investigation of their local elastic properties~\cite{Widmer-Cooper2008}.

We further remark that the defect picture does not rely on the introduction of defects of a specific size but rather on the existence of a characteristic size, $\xi_d$. 
It is then of interest to consider the distribution of the defect sizes, $P(\xi$). 
This distribution can be obtained from the distribution in frequency of the modes, given the QLMs ``dispersion relation'' $\xi = \xi(\omega)$, one might infer from Fig.~\ref{fig:comparexd}, as $D_{\rm loc}(\omega)d\omega = P(\xi)d\xi$.

The relation between boson peak, sound attenuation, and QLMs we have established calls for reconsidering previous works relating the elastic properties of amorphous materials to those of disorder mass-spring networks, e.g. see for a review~\cite{Nie2017}.
While elastic disorder induces a boson peak, our results suggest that only elastic networks with a disorder engineered to reproduce the observed connection between boson peak, localized modes and sound attenuation are relevant models for amorphous materials.

\begin{acknowledgments}
We thank E. Flenner and the authors of Refs.~\cite{wang2019sound,Shakerpoor2020} for kindly sharing their data, we used to produce
Fig.~\ref{fig:attPar}(b). 
We further thank W. Schirmacher for suggestions, and E. Lerner and K. Gonz\'alez-L\'opez for critical comments on a earlier version of the manuscript.
We acknowledge support from the Singapore Ministry of Education through the Singapore Academic Research Fund (MOE2017-T2-1-066 and MOE2019-T1-001-03), and the National Supercomputing Centre Singapore (NSCC) for the computational resources.
\end{acknowledgments}

\bibliography{Maintext}

\newpage
\setcounter{figure}{0}
\setcounter{equation}{0}
\newcommand{\sFrac}[2]{{\textstyle\frac{#1}{#2}}}
\def\u0#1{\underline {#1}}
\def\theequation{S\arabic{equation}}
\renewcommand*{\thefigure}{S\arabic{figure}}


\section{Disorder parameter and elastic correlation length}
According to corr-FET, the local shear modulus is correlated over an elastic length scale, $\xe$, and the fluctuations of the shear modulus coarse grained over regions of linear size $w$ containing $N$ particles, $\sigma^2_N$, asymptotically scale with $N\propto w^3$ as
\begin{equation}
    \frac{\sigma^2_N}{\mu^2} \propto \frac{\gamma}{w^3}
\end{equation}
where $\gamma$ is the disorder parameter. 
If a single length scale controls the emergence of this scaling regime, then
\begin{equation}
    \frac{\sigma^2}{\mu^2} = \left(\frac{\xe}{w}\right)^3 g\left(\frac{w}{\xe}\right)
\end{equation}
with $g(x)$ constant for $x = 1$.
This relation implies $\gamma \propto \xe^3$.

We have verified this relation and determined $\gamma$ using two approaches.
In the first case, we focused on the investigation of the sample-to-sample fluctuations of the elastic properties.
In the second one, we consider the fluctuations of the elastic properties defined over a coarse-graining length scale $w$.

\subsection{Sample-to-sample fluctuations}
\begin{figure}[!t]
 \centering
 \includegraphics[angle=0,width=0.45\textwidth]{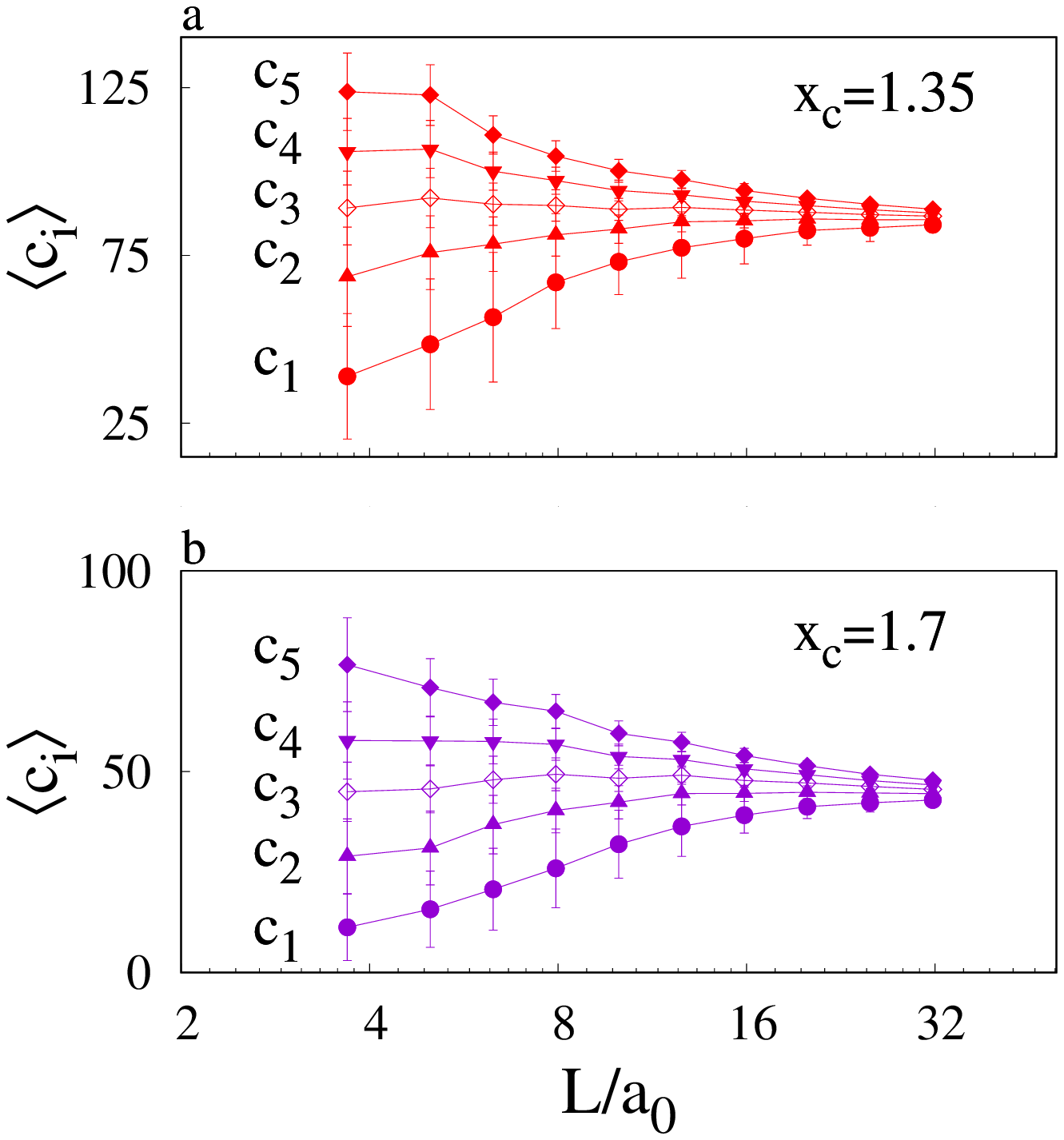}
 \caption{
System size dependence of sample-averaged eigenvalues $c_1 \leq \ldots \leq c_5$ for $x_{c} = 1.3$ (a) and $x_{c} = 1.6$ (b). Data are averaged over $200$ realizations for each system size and cutoff.
\label{fig:eigs}
}
\end{figure}

We have determined FET's disorder parameter, $\gamma$, and the elastic correlation length, $\xi_e$, investigating the size dependence of the fluctuations of the elastic response.
We characterize the elastic response of our samples evaluating~\cite{Tsamados2009, mahajan2021emergence} their stiffness matrix, $\hat {\bf C}$. 
We determine this matrix by measuring the stress changes resulting from six independent strain deformations followed by energy minimization in the linear response regime.
In three dimensions, five of eigenvalues of $\hat {\bf C}$, we indicate with $c_1 \leq \ldots \leq c_5$, should equal to $2\mu$. 
In Fig.~\ref{fig:eigs} we show that these eigenvalues approach a common limiting value as the linear size of the system $L$ increases.
Results in the two panels refer to two different values of the parameter $x_c$ fixing the width of the attractive well of the potential.

We study the fluctuations in the shear response ~\cite{mahajan2021emergence} investigating the size dependence of the normalized fluctuations
$
\frac{\sigma_c^2}{\langle c_i \rangle^2} = \frac{\overline{\langle c_i^2 \rangle} - \overline{\langle c_i \rangle^2}}{(2\mu)^2},
$
where $\overline{\cdot}$ indicate averages over the five eigenvalues, $\langle \cdot \rangle$ the sample averages, and $2\mu = \overline{\langle c_i\rangle}$.

\begin{figure}[!h]
 \centering
 \includegraphics[angle=0,width=0.45\textwidth]{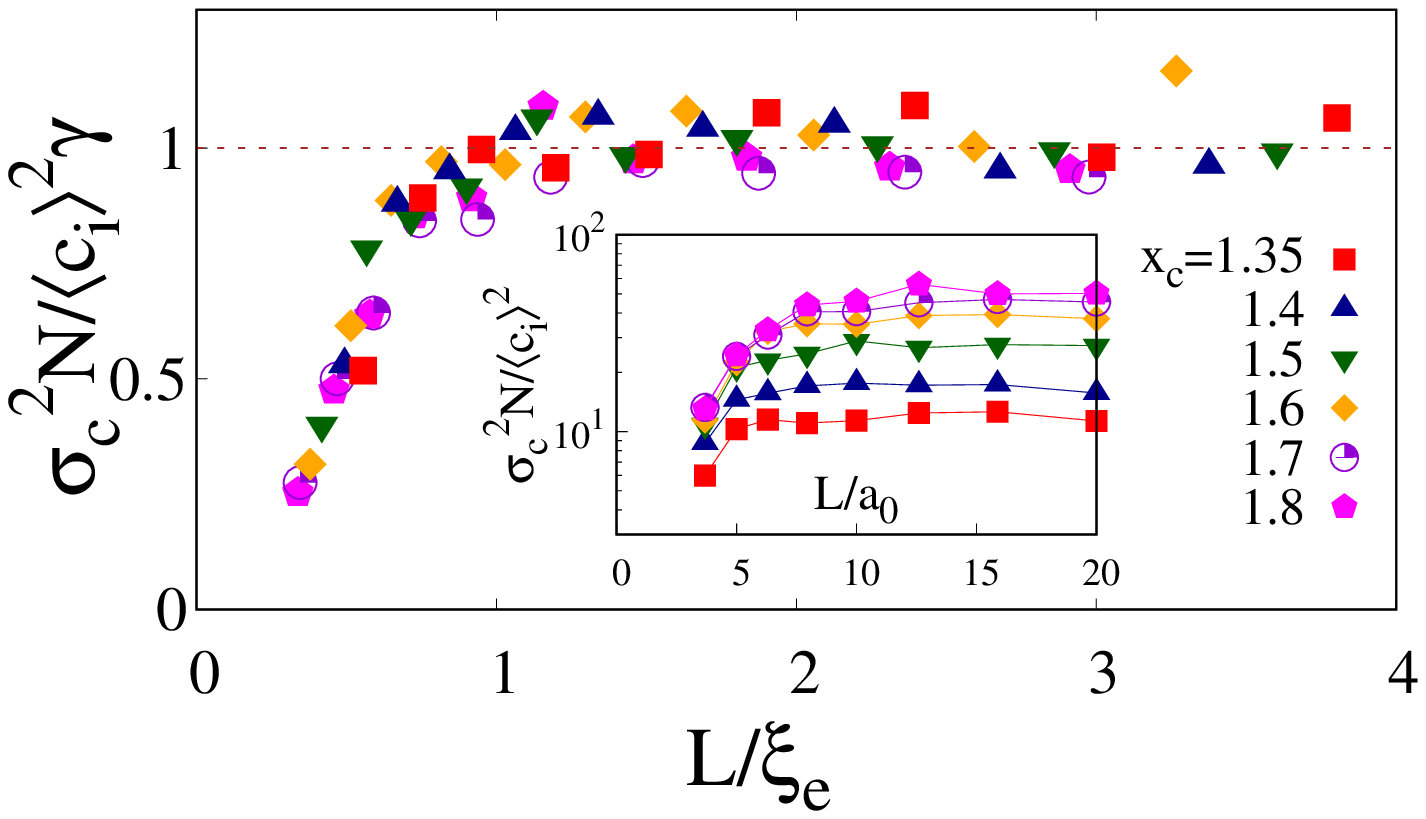}
 \caption{
 The fluctuations of the smallest eigenvalues of the stiffness matrix (inset) of systems of size $L$ collapse when their magnitude is scaled by $\gamma$, and $L$ is scaled by $\xe \simeq 3 a_0 \gamma^{1/3}$ (main panel). Panel b illustrates the $x_c$ dependence of $\xe$ and of $\gamma$.
\label{fig:gamma}
}
\end{figure}

Fig.~\ref{fig:gamma} (inset) show that the fluctuations of these five eigenvalues asymptotically as $L^{-3}$.
In the main panel, we scale-collapse data corresponding to different $x_c$ plotting them versus $L/\xe$ and normalizing then by $\gamma$.
Specifically, we have fixed $\gamma$ and $\xe \simeq 3 a_0\gamma^{1/3}$ so that the fluctuations approach one at $L \simeq \xe$. The $x_c$ dependence of $\gamma$ and $\xe$ is in Fig. 1 of the main text.

\subsection{Coarse-grained elastic properties}
We have alternatively determined $\gamma$ and $\xe$ from a study of the fluctuations of the elastic properties coarse-grained over different length scales.

To this end, we define the stress tensor at the scale of the single particle. 
Specifically, the stress tensor of particle $i$ is
\begin{equation}
\sigma_{\alpha\beta}^{(i)} = \frac{\rho}{2}\sum_j^{(i)} (r_\alpha f_\beta)_j
\end{equation}
where the sum is over all interaction forces involving particle $i$, and $\alpha,\beta \in \{x,y,z\}$. 
This definition, which assumes that all particles occupy the same volume, ensures that by summing the stress associated to the individual particles one recovers the macroscopic stress tensor.
We further define the single-particle stiffness-matrix as 
\begin{equation}
c^{(i)}_{ab} = \frac{d\sigma_{a}^{(i)}}{d\epsilon_{b}},
\end{equation}
where $\epsilon_{ab}$ is the macroscopic strain~\cite{Tong2020}, and $d\sigma_{a}^{(i)}$ the change in stress occurring after the system is deformed and its energy minimized. 
That is, our definition of stiffness-matrix accounts for the non-affine contributions to the elastic response.
The suffix $a,b=1,\ldots,6$ indicates $xx,yy,zz,xy,xz,yz$ so that, e.g., $c_{14}^{(i)}$ stands for $c_{xxxy}^{(i)}$. 
We remark that this is one among the possible definition of local stiffness tensor~\cite{Mizuno2013}.

We associate to each particle a shear modulus $g^{(i)} = c^{(i)}_{44}/2$.
The shear modulus of a N particle system, or equivalently of a region containing N particles, is 
$G_N = \frac{1}{N} \sum g^{(i)}$, where the sum is over the shear modulus of the $N$ particles. 

To investigate the dependence of the fluctuations $\sigma^2_w$ of the coarse grained shear modulus on the coarse graining length, $w$, we divide the simulation domain in $(L/w)^3$ cubic boxes. 
We define a shear modulus coarse grained over the length scale $w$ by associating to each box the average shear modulus of the particles whose centre lie in the box.

\begin{figure}[!t]
 \centering
 \includegraphics[angle=0,width=0.48\textwidth]{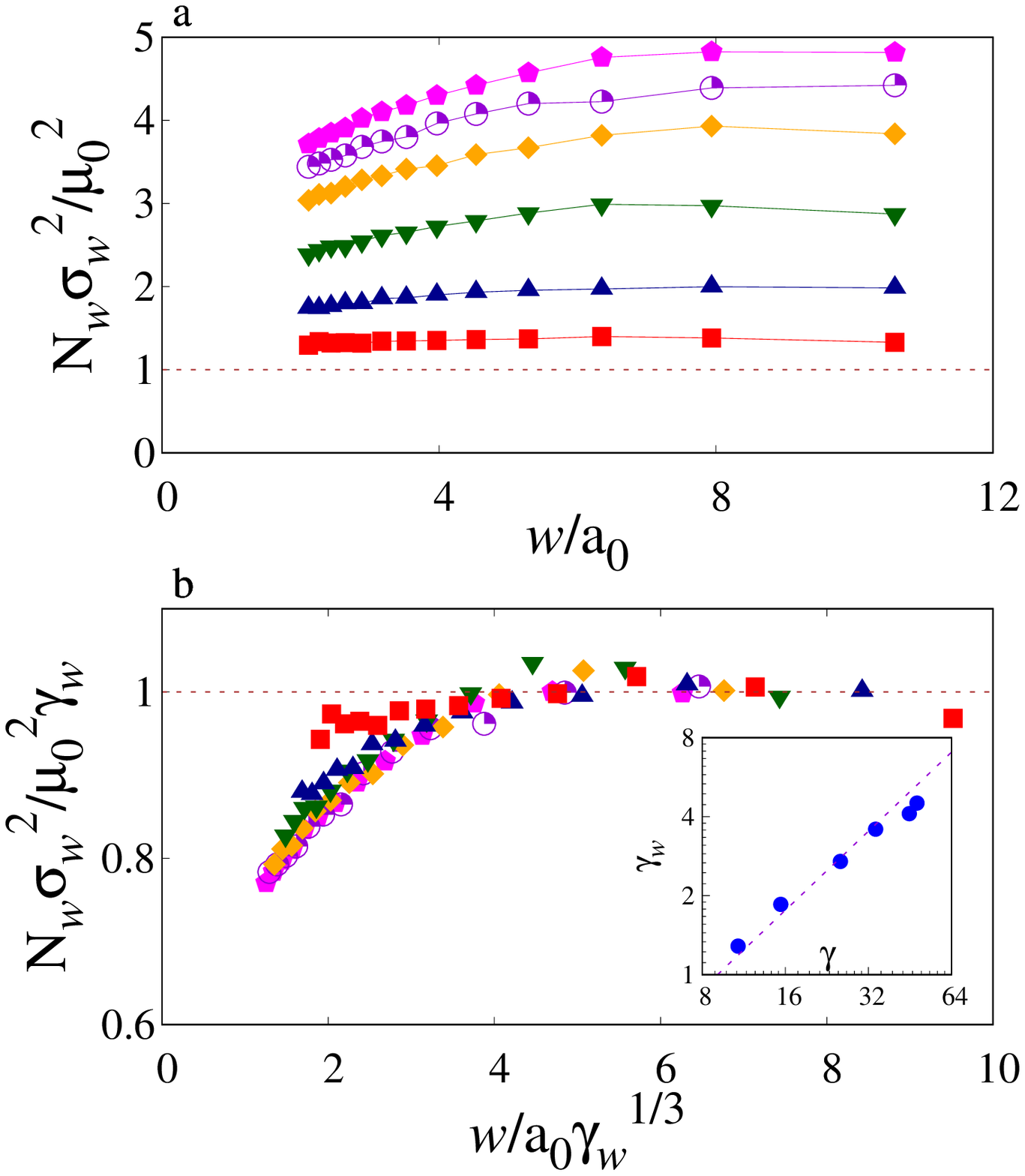}
 \caption{
 The fluctuations of the shear modulus of systems coarse-grained over length scale $w$ (a) collapse when their magnitude is scaled by $\gamma_w$, and $L$ is scaled by $\gamma_w^{1/3}$ (b).
 The inset of (b) shows that $\gamma_w$ is proportional to the disorder parameter $\gamma$ (Fig. 1b of the main text) determined via the study of the sample-to-sample fluctuations of the elastic properties.
 \label{fig:coarseFluc}
}
\end{figure}

We have determined the fluctuations $\sigma^2_w$ of the coarse-grained shear modulus for each interaction potential, $x_c$. 
Results are averaged over 200 independent realizations.
By central limit theorem, we expect these fluctuations to scale as the number of particles in the boxes, $\sigma^2_w \propto N_w\propto w^{-3}$.
Indeed, the normalized fluctuations $N_w \sigma^2_w/\mu_0^2$, with $\mu_0 = \langle \mu \rangle$, approach an $x_c$ dependent constant at large $w$, as we show in Fig.~\ref{fig:coarseFluc}a.

In Fig.~\ref{fig:coarseFluc}b, we show that these data collapse on a master curve if the fluctuations are scaled by $\gamma_w(x_c)$, which we fix so that the asymptotic value becomes one, and distances are scaled by $\gamma_w^{1/3}$. 
A small deviation only occurs at the smallest $x_c$, presumably because the elastic correlation length becomes comparable to the typical interparticle spacing. 
The value $\gamma_w$ leading to the data collapse is proportional to the disorder parameter $\gamma$ estimated from the sample-to-sample fluctuations of the local elastic properties, as we show in the inset.
Hence, these study demonstrate that $\gamma$ and $\xe$ can be equivalently determined investigating the size dependence of the sample-to-sample fluctuations of the elastic properties, or the size dependence of fluctuations of coarse-grained elastic properties.

\section{FET disorder parameter in the defect picture}
In the defect picture, we describe the material as an elastic continuum with shear modulus $\mu_0$, punctuated by $n$ spatially uncorrelated defects per unit volume, each defect being a region of volume $V_d \propto \xi_d$ with shear modulus $\mu_0+\delta\mu_d$.
A particle has shear modulus $\mu_0$ [$\mu_0+\delta\mu_d$] with probability $1-p$ [$p = nV_d = n \xi_d^3$].
In this picture,
\begin{eqnarray}
    \langle \mu \rangle &=& \mu_0 + p \delta\mu_d \\
\langle \mu^2 \rangle &=& 
\mu_0^2 + p \sigma_d^2 +2p\mu_0\sigma_d
\end{eqnarray}
If the density of defects is small, $p = nV_d\ll 1$, $\langle \mu \rangle \simeq  \mu_0$, and the variance of the single particle shear modulus is
\begin{equation}
\sigma_1^2 =  p(1-p)\delta\mu_d^2 \simeq p\delta\mu_d^2.
\end{equation}
By central limit theorem the fluctuations of the shear modulus of regions containing $N$ particles scale asymptotically as 
\begin{equation}
\frac{\sigma^2_N}{\mu_0^2} 
\propto 
\frac{1}{\mu_0^2}\frac{\sigma^2_1}{N} \propto 
\frac{1}{\mu_0^2}
\frac{p\delta\mu_d^2}{N} \propto (na_0^3) \left( \frac{\xi_d}{a_0}\right)^3 \frac{\delta\mu_d^2}{\mu_0^2}\frac{1}{N}
\end{equation}
and FET disorder parameter results
\begin{equation}
\gamma \propto (na_0^3) \left( \frac{\xi_d}{a_0}\right)^3 \frac{\delta\mu_d^2}{\mu_0^2}\
\end{equation}

\end{document}